# Multiplexed wavefront sensing with a thin diffuser

TENGFEI WU,[1,2] MARC GUILLON,[2,3] GILLES TESSIER,[1] PASCAL BERTO[1,2*]

[1]Sorbonne Université, CNRS, INSERM, Institut de la Vision, 17 Rue Moreau, 75012 Paris, France
[2]Université de Paris, SPPIN – Saints-Pères Paris Institute for the Neurosciences, CNRS, 75006 Paris, France
[3]Institut Universitaire de France (IUF), Paris, France
*Corresponding author: pascal.berto@u-paris.fr

**In astronomy or biological imaging, refractive index inhomogeneities of e.g. atmosphere or tissues induce optical aberrations which degrade the desired information hidden behind the medium. A standard approach consists in measuring these aberrations with a wavefront sensor (e.g Shack-Hartmann) located in the pupil plane, and compensating them either digitally or by adaptive optics with a wavefront shaper. However, in its usual implementation this strategy can only extract aberrations within a single isoplanatic patch, i.e. a region where the aberrations remain correlated. This limitation severely reduces the effective field-of-view in which the correction can be performed. Here, we propose a new wavefront sensing method capable of measuring, in a single shot, various pupil aberrations corresponding to multiple isoplanatic patches. The method, based on a thin diffuser (i.e a random phase mask), exploits the dissimilarity between different speckle regions to multiplex several wavefronts incoming from various incidence angles. We present proof-of-concept experiments carried out in wide-field fluorescence microscopy. A digital deconvolution procedure in each isoplanatic patch yields accurate aberration correction within an extended field-of-view. This approach is of interest for adaptive optics applications as well as diffractive optical tomography.**

## 1. INTRODUCTION

Image degradation caused by aberrations can be a critical issue when imaging objects of interest located behind a complex medium, e.g. a turbulent atmosphere or biological tissues, in which the distribution of the refractive index is both heterogeneous and dynamic. To correct aberrations and increase the image contrast and resolution, the most common technique is to use pupil adaptive optics (AO), in which the aberration is first characterized in the pupil plane of the optical system, and then corrected either digitally or by a wavefront shaper [1–4].

Aberration characterization methods can be classified into either direct or indirect wavefront sensing techniques. Indirect techniques estimate aberrations by optimizing, over time, a feedback image or signal with certain metrics (e.g. image contrast, sharpness or intensity [5–7], nonlinear signals [8–10] etc.). These techniques, also referred to as "sensorless", do not require any dedicated optical device to estimate aberrations, thus potentially reducing cost and hardware complexity. However, the convergence of the optimization process can be unstable and/or may lead to local minima. More importantly, these techniques require multiple measurements, which limits their application to static, or slowly varying, aberrating media, thus excluding most astronomical and ophthalmologic applications.

The other class of approaches, direct wavefront sensing, relies on a deterministic measurement of aberrations, typically using point sources called guide-stars (GS). The most common wavefront sensor (WFS) is indisputably the Shack Hartman [11]. This WFS generally samples the pupil plane with a micro-lens array, where each micro-lens directly yields the local phase gradient of the wavefront in the corresponding subset of camera pixels, which defines a macropixel. A significant advantage of this method is that it only requires a single image acquisition to measure a wavefront, thus allowing higher temporal resolution. Direct wavefront sensing is preferred – and often crucial – whenever dynamic aberrations are involved, like those caused by the atmospheric turbulence in astronomy or the lachrymal film in ophthalmology [4,12].

However, aberrations only remain correlated within a limited region called the isoplanatic patch, whose typical size depends on the position and properties of the aberrating medium. A pupil aberration measurement typically assumes a 2D projection of the 3D refractive index of this medium along the direction of the GS. Wavefront measurements obtained with a single GS can therefore only be used to correct the aberration within a limited field-of-view (FoV) in the case of volumetric aberrating media. Aberration correction is then only valid in the angular vicinity of the GS, but decorrelates away from it. Extensive efforts have been made to address this critical issue using direct or indirect wavefront sensing methods. Conjugate AO can increase the isoplanatic patch size by conjugating the WFS (and the corrector) to the main aberrating layer [13–16]. This approach is therefore highly efficient to widen the FoV in situations where aberrations originate from a single, well-defined planar layer. However, it requires a much larger number of sensing/shaping modes and it is not optimal in the most common scenarios where aberrations originate from a non-planar interface, from several layers, or from a volume. Ideally, a tomographic characterization of the aberrating medium is needed to retrieve high resolution over a large FoV or volume. This concept is at the heart of Multi-Conjugate AO, which uses several WFS (typ. >6) to probe aberrations along several GSs directions in order to estimate volumetric aberrations and correct them in several planes [17,18] . While increasingly used in astronomy, this approach is too complex, cumbersome and expensive to be routinely applied in microscopy and ophthalmology.

Some approaches operating in the pupil plane rather than in a conjugated plane have also emerged. An interesting method sequentially applies various masks in the pupil plane to retrieve the local phase gradient in multiple isoplanatic patches [19–21]. Alternatively, the beam or GS can be scanned to sequentially measure pupil aberrations in different locations [22]. Both approaches however require a time-consuming scan which degrades temporal resolution. Shack-Hartmann WFS could in principle provide parallelized multi-GS, multi-angle measurements in a single shot but their angular dynamics is limited by aliasing problems and their ability to measure multiple, distributed GSs is strongly constrained (See Supplementary S1).

While some of these emerging methods increasingly allow aberration characterization along multiple angles, they can hardly address dynamic modifications. Conversely, fast methods are essentially limited to a single GS. In microscopy, the speed constraint is partially relieved because aberrations in tissues typically evolve over time scales of a few minutes. However, the development of fast, multiplexed wavefront sensing methods remains essential when imaging moving organisms, or deep inside tissues, where wavefronts are also affected by rapidly evolving, multiple scattering [23]. It is also crucial for the volumetric imaging (e.g. light sheet, light field, multiplane imaging …) of large samples since aberrations need to be characterized for a high number of isoplanatic volumes, which is time-consuming with indirect or sequential methods [24–27]. In most fields requiring aberration correction, a simultaneous parallel measurement of multiple pupil aberrations would provide a broad FoV. Such approaches could also put single shot tomographic microscopy within reach, while drastically simplifying the instrumentation.

In this context, recent advances in light manipulation through scattering media appear as decisive. The unique signature associated to random speckle patterns makes them particularly suited to the unambiguous retrieval of multiplexed information. Notably, the use of diffusers as encoding masks has recently shown its potential in areas including compressive ghost imaging [28], single-exposure 2D or 3D imaging [29–31], hyperspectral imaging [32–34] and 3D super-resolution microscopy [35]. Furthermore, the analysis of speckle pattern distortions has recently emerged as an accurate way to perform high-resolution wavefront sensing [36–39].

This paper addresses the single-shot multiplexed measurement of multiple pupil aberrations, i.e. the parallel measurement of several wavefronts coming from different propagation directions. To solve the assignment ambiguity problem inherent to multiple wavefronts and periodic mask patterns (e.g. Shack-Hartmann), we propose to use a thin diffuser (i.e. a random phase mask) as Hartmann mask. The wavefronts related to several GSs are thus encoded into an incoherent sum of distorted speckle patterns. By exploiting both the dissimilarity between different speckle regions and the large memory effect of a thin diffuser [40,41], we show that each speckle pattern can be unambiguously ascribed to its GS, allowing to extract the associated phase gradient maps from speckle pattern distortions. We experimentally validate this concept in microscopy with the multiplexed acquisition of wavefronts originating from several fluorescent GSs. This demonstrates the efficacy of this single shot method for the characterization of pupil aberrations in multiple isoplanatic patches. Finally, we show in a proof-of-concept experiment that this multiplexed WFS can feed a deconvolution procedure, to digitally correct aberrations in several isoplanatic patches and obtain high-resolution images in a large FoV.

## 2. PRINCIPLE

Figure 1(a) illustrates the general concept of multiplexed wavefront sensing in the context of fluorescence microscopy. Fluorescent GSs are imaged by a microscope objective, through an aberrating medium. The latter induces spatially varying pupil aberrations as the GSs are located in different isoplanatic patches. In the pupil plane, each GS will then generate an aberrating wavefront carrying a global tilt offset according to the GS lateral position. A thin-diffuser-based WFS, conjugated to the pupil plane, then allows measuring these aberrating wavefronts incoming with different propagation directions from a single image acquisition.

Before detailing the multiplexed wavefront sensing concept, let's first recall the basics of diffuser-based WFS in the case of a single wavefront acquisition. We recently demonstrated that a thin diffuser can be used as a Hartmann mask to perform broadband wavefront sensing with a high accuracy and resolution [39,42,43]. The WFS is composed of a thin diffuser used as a Hartmann-type mask located at a short distance $d$ from a camera sensor [see Fig. 1(b)]. A "reference" speckle pattern $R(\boldsymbol{r})$ is first acquired by illuminating the WFS with a plane wave. Due to the large angular memory effect of thin diffusers [40,41], a tilt to the wavefront does not modify or decorrelate the speckle pattern but simply translates it. For a distorted wavefront, speckle grains will be locally shifted on the detector according to the local wavefront gradient, as shown in Fig. 1(b): a speckle grain at position $\boldsymbol{r}$ is locally displaced by a vector $\boldsymbol{u}(\boldsymbol{r})$, and the resulting speckle map $S(\boldsymbol{r})$ is distorted as compared to the reference $R(\boldsymbol{r})$ according to [36,43]:

$$S(\boldsymbol{r}) = I(\boldsymbol{r})R[\boldsymbol{r} - \boldsymbol{u}(\boldsymbol{r})] \qquad (1)$$

where $I(\boldsymbol{r})$ is the normalized intensity map. The phase gradient map $\boldsymbol{\nabla}_\perp \varphi$ can then be estimated by measuring the overall speckle grain displacement map $\boldsymbol{u}$ [39]:

$$\boldsymbol{\nabla}_\perp \varphi \simeq k_0 \boldsymbol{u}/d \qquad (2)$$

with $k_0$ the wavenumber.

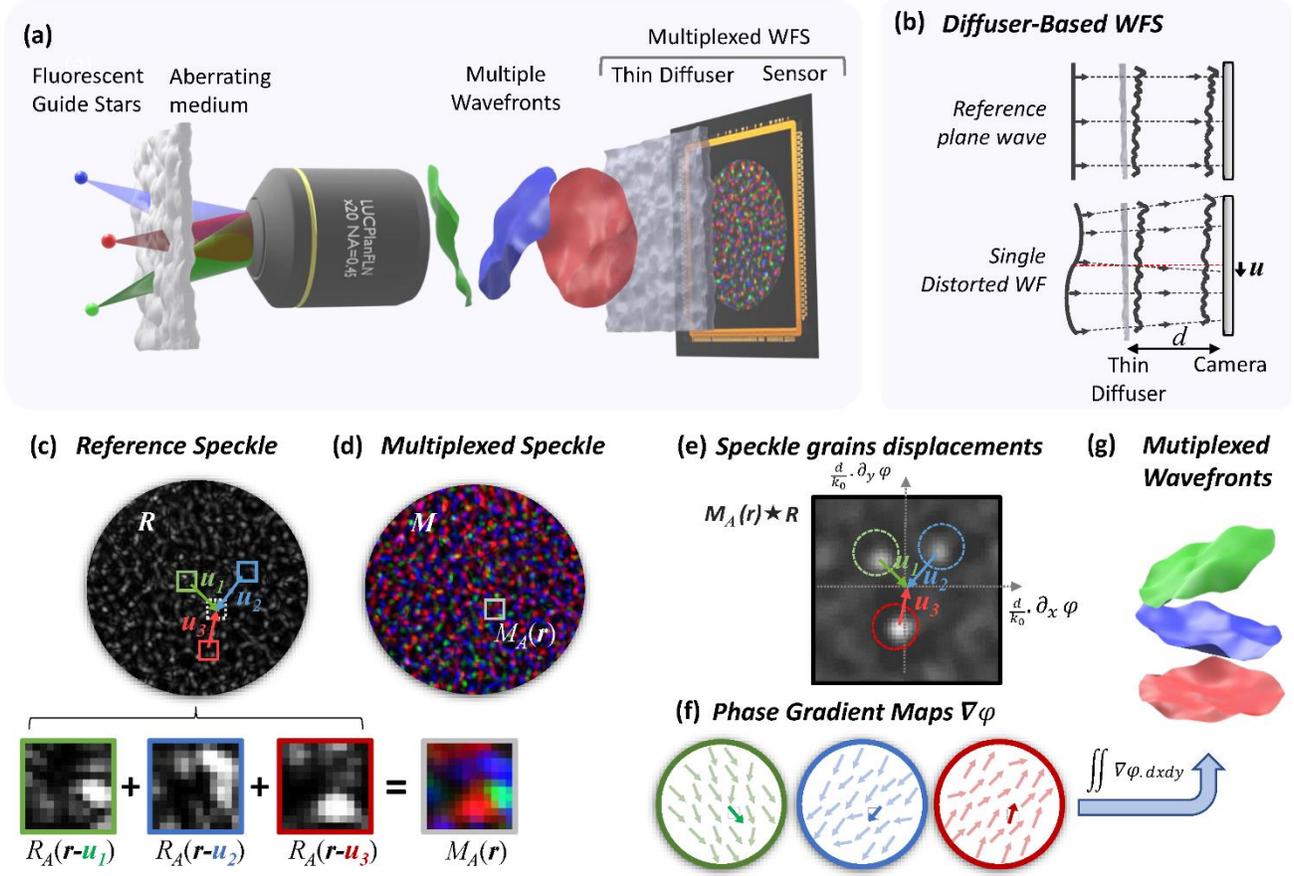

**Fig. 1**. Principle of angularly multiplexed wavefront sensing. (a) Schematic description of the concept, here applied to fluorescence microscopy. $N=3$ fluorescent guide stars are imaged through an aberrating layer by a microscope objective. This leads to $N=3$ tilted and aberrated wavefronts in its pupil plane. A wavefront sensor (WFS) based on a thin diffuser allows to (g) multiplex the acquisition of these wavefronts. (b) Basic working principle of the thin diffuser-based WFS: A plane wave illuminating a 1° holographic diffuser generates a reference speckle pattern $R$ on a camera sensor located at a distance $d$. For a distorted wavefront, measurement of the speckle grain displacements $\boldsymbol{u}$ give access to the phase gradient ($\boldsymbol{\nabla}_\perp \varphi \simeq k_0 \boldsymbol{u}/d$). (c,d) The reference speckle pattern $R$ and the multiplexed speckle pattern $M$ measured by the gray-level camera contain information on the three wavefronts. $M$ is the incoherent sum of three distorted speckle patterns related to each wavefront. The pattern measured within a chosen region, the macropixel $M_A$ (gray square in d.), can be described as the sum of three patterns of the reference (green, blue and red boxes in c.) each shifted by $\boldsymbol{u}_j$, with $j=1,2,3$, the index of each wavefront. (e) Owing to the orthogonality between different speckle regions, the cross-correlation of $M_A(\boldsymbol{r})$ with the full reference speckle pattern $R$ reveals $N=3$ peaks. The peak position gives access to the local speckle grains displacements related to each wavefront, and so, to (f) the phase gradient maps of the wavefronts. (g) A 2D integration step finally allows independent wavefronts reconstructions.

Equation 2 shows that thin-diffuser-based WFS and microlens-array-based Shack Hartmann WFS have similar behaviors, with speckle grains playing a role similar to that of focal spots in each subset of camera pixels, or macropixel (typ. 7x7 camera pixels [39]), and the refractive diffuser surface behaving as random, aberrating microlenses [29]. Interestingly, the use of a thin diffuser fundamentally provides a major advantage in the context of wavefront multiplexing. While the intensity pattern generated by micro-lens arrays in a Shack Hartmann is periodic, and thus prone to ambiguity between identical, unrecognizable spots, a diffuser generates a superposition of unique patterns which can be identified unambiguously. The ambiguity problem strongly limits the phase dynamic range and the number of wavefronts that can be multiplexed with a Shack Hartmann (See Supplementary S1). In contrast, random speckle grains provide a unique signature. Two uncorrelated, random speckle patterns are indeed statistically orthogonal relative to a zero-mean cross-correlation product. This inherent property of speckles alleviates the ambiguity issue and allows to retrieve multiple wavefronts locally encoded in the form of multiple, orthogonal speckle patterns [35].

To illustrate this concept, Fig. 1(c) and (d) respectively show a reference speckle pattern $R$ and a multiplexed speckle pattern $M$ encoding three wavefronts. Note that although the three speckle patterns associated to each wavefront are represented using three colors for clarity, the concept clearly applies to a single wavelength, monochrome detector, and to more than $N=3$ wavefronts. Since the $N$ wavefronts originate from different incoherent point sources, the pattern $M$ can be described as the sum of $N$ reference speckle patterns $R$, each shifted and deformed by $N$ different non-rigid transformations $\boldsymbol{u}_j(\boldsymbol{r})$:

$$M(\boldsymbol{r}) = \sum_{j=1}^{N} S_j(\boldsymbol{r}) = \sum_{j=1}^{N} I_j(\boldsymbol{r}) R[\boldsymbol{r} - \boldsymbol{u}_j(\boldsymbol{r})] \qquad (3)$$

where $j$ stands for the index of the wavefront among the $N$ multiplexed ones ($N=3$ in Fig. 1). This superposition is illustrated for a chosen subset of $M(\boldsymbol{r})$, the macropixel $M_A(\boldsymbol{r})$ (here e.g. 12x12 pixels) shown in Fig 1(c) and (d): the intensity in the macropixel

$M_A(r)$ is the incoherent sum of three different subregions of the reference speckle $R$ having undergone different lateral shifts, $u_1$, $u_2$ and $u_3$ (See zoomed macropixel in Fig. 1(c) and (d)). Since these subregions correspond to different regions of the speckle pattern created by different parts of the diffuser Hartmann mask, they are statistically orthogonal. The cross-correlation $M_A(r) \star R$ therefore allows to extract, without ambiguity, the speckle grain displacements vector map $u_j(r)$ associated to each wavefront through an estimate (centroid) of the position of the peaks, and the normalized intensities $I_j$ through the maximum values in each peak. This process is illustrated in Fig. 1(e), where three correlation peaks yielding $u_1$, $u_2$ and $u_3$ are clearly visible. The local phase gradient at the pupil coordinate $r$ can then be estimated using Eq. (2). Using a Digital Image Correlation (DIC) algorithm [44], the same process can be reproduced in all macropixels $M_A(r')$ of the full multiplexed speckle pattern $M$ in order to extract the phase gradient vector maps associated to each wavefront [see Fig. 1(f)]. Finally, a 2D integration of these phase gradient maps can be used to retrieve the $N$ wavefronts [see Fig. 1(g)].

It should be noted that this method implicitly requires a sparsity assumption about GSs density which is clearly visible in Fig 1(e): the position of the correlation peaks, related to the maximum phase gradients of wavefronts (delimited here by dashed circles), can only be estimated if the correlation peaks do not overlap (See Supplementary S2). However, this constraint is common to all WFS, and is not critical to the targeted application since wavefronts coming from GS located in different isoplanatic patches are generally well separated angularly. Furthermore, as discussed in the experimental section, the validity of this assumption can be predicted by calculating the "global" cross-correlation $M \star R$.

This approach can in principle be applied to any number of GS and wavefronts. However, the contrast of the multiplexed speckle pattern $M(r)$ resulting from the superposition of $N$ patterns ($C \propto 1/\sqrt{N}$) decreases for large $N$-value. This reduces the signal to noise ratio of the cross-correlation map $M_A(r) \star R$, the accuracy of the determination of speckle grain displacement $u_j$, and thus the accuracy of wavefront reconstruction. A first solution to maintain a high reconstruction fidelity while multiplexing a large number $N$ of wavefronts would be to use larger macropixels $M_A$ (e.g. 45x45 pixels, Supplementary S3). However, this method drastically degrades the spatial resolution on the rebuilt wavefront. To alleviate resolution degradation, we propose an iterative DIC process to converge towards each speckle pattern $S_k$ [See Eq. (4)]: after a first DIC step (which provides a first estimation of the intensities $I_j$ and displacement maps $u_j$), the distorted speckle pattern $S_k$ can be isolated from the multiplexed speckle pattern $M$ by subtracting the contributions of all others GSs $j \neq k$ :

$$S_k(r) = M(r) - \sum_{j \neq k}^N I_j(r) R[r - u_j(r)] \quad (4)$$

This process restores the contrast of speckle pattern $S_k$ as compared to $M$ and thus allows, through a second DIC step, to recover a more accurate estimation of the intensities $I_k$ and displacement maps $u_k$. Noteworthy, since the subtracted patterns are prone to uncertainties, the contrast of the pattern $S_k$ is restored but remains noisy before the algorithm converges. For this reason, this process can be repeated T times to increase the reconstruction accuracy and can be seen as a gradient descent algorithm which iteratively minimizes the quantity $\left| M(r) - \sum_{j=1}^N I_j(r) R[r - u_j(r)] \right|^2$. A detailed description of the corresponding algorithm, as well as numerical simulation results demonstrating its relevance for high-resolution wavefront sensing are provided in Supplementary S3. Briefly, these simulations show that the number T of iterations which are necessary to retrieve $N$ wavefronts increases with $N$ (typ. T=2 for $N$=5 wavefronts). Supplementary S3 also shows that more than $N$=16 wavefronts can be retrieved using this iterative DIC approach, while reducing the RMS error by an order of magnitude compared to direct DIC (See Fig. S6.). Importantly, it also shows that high-resolution is preserved by this method since small phase pixel size (typ. 7x7 pixels) can be reached. When using a 4.2M pixel camera, this provides typ. 16 multiplexed wavefronts, each with 85K phase and intensity pixels.

## 3. RESULTS

### 3.1. Description of the setup

To validate this concept experimentally, we built a wide-field fluorescence microscope based on a commercial inverted microscope (Olympus IX-71). 1μm-diameter fluorescent beads randomly distributed on a glass slide are used as sample (Orange 540/560, Thermo Fisher). An aberrating layer (1° Holographic Diffusers, Edmund Optics) is positioned 150 μm away from the microbeads sample to induce spatially varying pupil aberrations [see Fig. 1(a)]. In order to excite the fluorescence of multiple beads chosen within the FoV, a 532nm laser beam is shaped using a phase-only spatial light modulator (SLM, X13138-01, Hamamatsu) conjugated to the back focal plane of the microscope objective (LUCPlanFLN, NA=0.45, x20, Olympus). The computer-generated hologram displayed on this SLM is calculated using a Gerchberg-Saxton algorithm [45] so as to illuminate the chosen beads and use them as GSs. The 532 nm excitation is spectrally filtered by a dichroic mirror (NFD01-532, Semrock) and a notch filter ($\lambda$=533 ± 2nm, Thorlabs) to collect the emitted fluorescence (See Supplementary S4).

The WFS is composed of a thin diffuser (1° scattering angle holographic diffuser, Edmund Optics) and a sCMOS camera (Zyla 5.5, Andor). The diffuser-camera distance is set to $d$=3 mm (see Ref. [39]), but placing the diffuser so close to the sensor was not mechanically possible. A 1x magnification relay lens (not shown in Fig. 1(a)) is therefore used to image the diffuser at a distance $d$ from the camera. To properly measure wavefront distortions coming from various GSs, the multiplexed WFS is conjugated with the back pupil plane of the microscope objective. The reference speckle pattern is acquired in a first step using a simple collimated beam generated from a multimode fiber (core diameter 10μm, Thorlabs) and a long focal length lens (f'=400mm). This reference speckle pattern is shown in Fig. 2(a).

### 3.2. Multiplexed wavefront sensing validation

To demonstrate the possibilities of the method, we first excited simultaneously $N$=3 GSs located in different isoplanatic patches, i.e. separated by more than 120 μm in the sample plane (See Supplementary S4 for the characterization of the sample isoplanatic patch size). On the WFS, their fluorescence yields a superimposition of 3 speckle patterns, as shown in the bottom of Fig. 2(a). The cross-correlation $M \star R$ between the reference and multiplexed speckles is shown in Fig. 2(b), clearly revealing the number and location of the three excited GSs. Here, the position of each correlation peak indeed indicates the average propagation direction (or global tip/tilt: $\langle\alpha\rangle_j = \langle u\rangle_j/d$) corresponding to each GS. The absence of overlap between peaks ensures that the sparsity hypothesis is valid, and that the wavefronts can be reconstructed independently. To this aim, the iterative DIC algorithm (T=3 iterations) is used to recover the speckle grains displacement maps related to each wavefront. The wavefronts obtained after integration of the phase gradient maps are shown in Fig. 2(c).

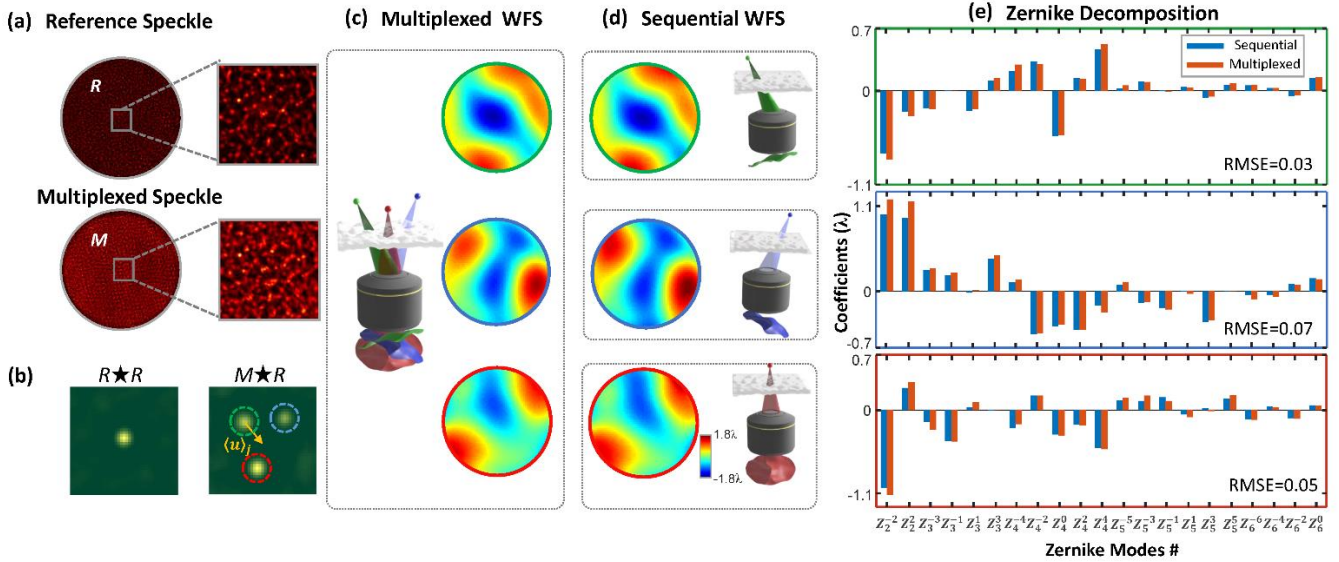

**Fig. 2. Experimental validation: single-shot measurements of three multiplexed wavefronts.** (a) Acquired reference speckle pattern R (for a plane wave illumination) and multiplexed speckle pattern $M$. Inset: zoom showing a reduced contrast for the multiplexed case due to the incoherent summation of speckle patterns. (b) Autocorrelation of the reference speckle pattern $R$, and cross-correlation of the multiplexed speckle pattern $M$ with the reference $R$. The three peaks in the cross-correlation reveal the number of multiplexed wavefronts and their global tip/tilt $\langle\alpha\rangle_j$ which can be determined by measuring the peaks shift ($\langle u\rangle_j = \langle\alpha\rangle_j d$) from the center. (c) Local cross correlation allows to reconstruct the three multiplexed wavefronts. Here, the tip/tilt of each wavefront has been subtracted. (d) Comparison with the "classical" sequential method and (e) Zernike decomposition (first 25 modes without piston, tip/tilt) for the three wavefronts. The excellent agreement between both measurements validates the multiplexing method.

The validity of these multiplexed measurements was compared to individual, sequential measurements with single GSs, as quantitatively validated in Ref. [39]. The multiplexed [Fig. 2(c)] and sequential [Fig. 2(d)] aberration measurements appear in excellent agreement. To quantify this comparison, we performed a Zernike decomposition on both acquisitions. In Fig. 2(e), we compare the first 25 lowest order modes. Note that tip/tilt ($Z_1^{-1}$ and $Z_1^1$) and defocus ($Z_2^0$) are omitted here for clarity because they respectively dominate other modes by more than one order of magnitude. Moreover, defocus strongly depends on the residual divergence of the reference plane wave and is therefore non-significant. As can be seen in Fig. 3(e), the agreement between sequential and multiplexed measurements is excellent (RMSE < 0.06λ), showing that several wavefronts generated by multiple GSs undergoing different aberrations can indeed be measured simultaneously in the pupil. In Supplementary S5, we present another experiment where $N$=5 GSs are multiplexed. We also discuss the experimental gain brought by the iterative DIC approach compared to the direct approach (RMSE reduction by a factor 1.5 to 6).

### 3.3. Correction beyond the isoplanatic patch

Having validated the ability to characterize multiple wavefronts simultaneously, we propose a proof-of-concept experiment showing that the measured wavefronts can be used to correct spatially-varying pupil aberrations. For this purpose, a second sCMOS camera (Panda 4.2, PCO) conjugated to the sample plane by a 4-f system images the fluorescent sample (See Supplementary S4). Figures 3(a) and 3(b) show the images of the sample without and with an aberrating medium, respectively.

When imaging an incoherent object described by its brightness $f(s)$, the image $g(r)$ obtained in the presence of aberrations can be written as [46]:

$$g(r) = \int \kappa(r - s, s) f(s) ds \qquad (5)$$

where $r$ and $s$ are spatial coordinates and $\kappa(q, s)$ describes the point-spread function (PSF) for a point source at a position $s$ in the FoV. The image is therefore the incoherent sum of contributions affected by aberrations which vary spatially (or angularly).

In most approaches, however, the PSF is assumed to be stationary, i.e. spatially invariant. Under this approximation, $\kappa$ does not depend on the observation direction, and can therefore be simply written $\kappa(q, s) \approx \kappa(q)$ all across the field of view. The aberrated image described by Eq. (5) then becomes a simple convolution product:

$$g(r) = \int \kappa(r - s) f(s) ds = (\kappa \circledast f)(r) \qquad (6)$$

If the aberrated wavefront $P(q')$ is measured in the pupil plane using a given GS, the stationary PSF $\kappa(q)$ can be estimated using a Fourier transform: $\kappa(q) = |F\{P(q')\}|^2$, where $|F\{\cdot\}|^2$ is the square modulus of the Fourier transform. As shown by Eq. (6), a simple deconvolution of $g(r)$ by $\kappa(q)$, using a Richardson-Lucy algorithm (Matlab image processing Toolbox, 15 iterations) then yields a corrected image $f(r)$. This is rigorously accurate in the direction of the GS used to measure $P(q')$, and the correction remains acceptable in its vicinity (within the *isoplanatic patch*), but it quickly degrades away from it, beyond a distance $R_{patch}$. This is clearly visible in Fig. 3(c)), where this deconvolution process was applied using a distorted wavefront measured with a single GS (top right).

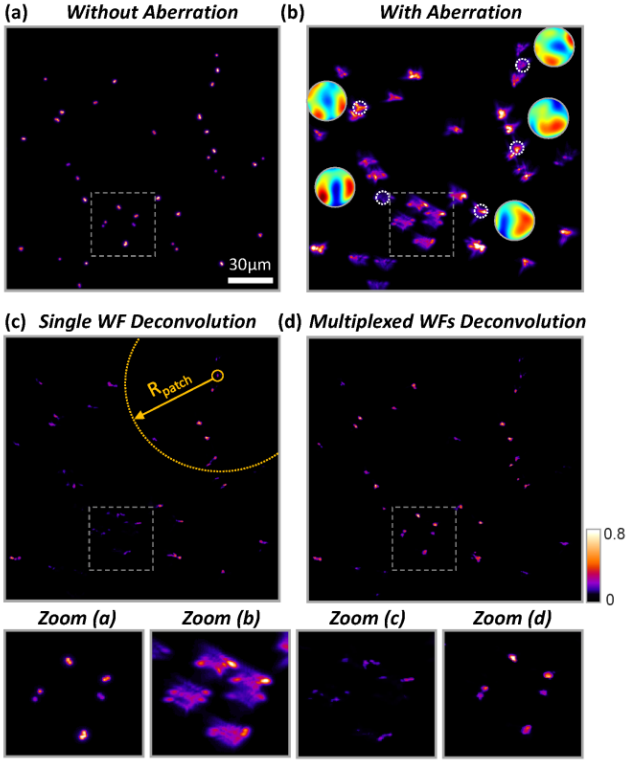

**Fig. 3. Aberrations deconvolution using multiplexed WFS** (proof-of-concept experiment). (a) Non-aberrated image of fluorescent beads used as ground truth. (b) Image acquired in presence of an aberrations medium (c) Aberration correction using a single wavefront measurement (from the GS indicated with a solid circle). The aberration is corrected properly within the isoplanatic patch (indicated by the dashed circle), while the correction fails in the other region. (d) Aberration correction with five wavefronts acquired in a single shot. Each of the multiplexed wavefront is used to correct a certain region within the corresponding isoplanatic patch and a proper image stitch enables to correct the aberrations over a larger FoV. A zoomed is also provided for the 4 cases.

In this context, the simultaneous measurement of wavefronts coming from multiple GSs described above can provide a better estimation of the angle-dependent PSF $\kappa(\boldsymbol{q}, \boldsymbol{s})$, particularly in cases where aberrations decorrelation over time demands simultaneous measurements across the FoV. In the proof-of-concept experiment shown in Fig. 3(d), the wavefronts coming from $N$=5 GSs, located in different isoplanatic patches, were simultaneously measured to estimate $P_j(\boldsymbol{q}')$ ($j$=1 to 5, corresponding wavefronts shown in Fig. 3(b) and deduce the associated PSF $\kappa_j(\boldsymbol{q})$ in each isoplanatic patch. The aberrated image can then be divided into $N$=5 regions delineated by indicative functions $\iota_j(\boldsymbol{r}) \in \{0,1\}$ which are equal to 1 inside the j$^{th}$ region and 0 elsewhere [46]:

$$g(\boldsymbol{r}) \approx \sum_{j=1}^{N=5} \iota_j(\boldsymbol{r})\big(\kappa_j \circledast f\big)(\boldsymbol{r}) \qquad (7)$$

The image shown in Fig. 3(d) was corrected using this piecewise approximation, i.e. by performing a deconvolution in each isoplanatic patch with the associated PSF $\kappa_j$. The improvement over the single, stationary WF correction $\kappa$ [Fig. 3(c)] is significant in the entire FoV, both in terms of resolution and contrast, providing image improvements beyond the isoplanatic patch (See also Supplementary S6). Since the $N$ measurements are performed simultaneously, the method is of particular interest for applications involving time-dependent aberrations.

## 4. CONCLUSION AND DISCUSSION

We proposed and demonstrated the use of a thin diffuser to simultaneously acquire multiple aberrated wavefronts by recording a single speckle pattern image. This direct, multi-angle wavefront sensing approach provides deterministic and quantitative measurements. It exploits the large memory effect of thin diffusers as well as the statistical orthogonality of speckle patterns to solve the aliasing problem intrinsic to periodic Hartmann masks. The proposed DIC-based algorithm can successfully reconstruct several (5 or more, see Supplementary S3) angularly-multiplexed wavefronts with high precision (typ. RMSE<$\lambda$/15) and high resolution (85K phase and intensity pixels), even for large angular distances between wavefronts. When conjugated to the pupil plane of an imaging system, this multi-angle WFS can thus sense, in a single shot, the aberrations of GSs located in different isoplanatic patches of the FoV. We illustrated the potential of this method for the digital correction of aberrations: several PSFs can be estimated simultaneously, and used to accurately deconvolve an aberrated image in multiple isoplanatic patches in order to recover a high-resolution image in the entire FoV. Here, the correction is performed considering a discrete set of GS and aberrated wavefronts, in the corresponding patches, but a better image correction could be obtained by interpolating between patches, and precisely estimating the PSF in each point of the FoV [46].

Multi-angle wavefront sensing has the potential to benefit a wide range of applications, in linear or non-linear microscopy. Coupled to light-sheet excitation, this instantaneous, large FoV digital AO method could be highly valuable to image freely moving animals (e.g swimming Zebrafish or C-Elegans) or large-volume samples [47]. Additionally, single-shot aberrations measurement in multiple patches (using multi-spot excitation) has an interesting potential to accelerate deep tissue imaging. In such cases, the isoplanatic patch size can be reduced to a few micrometers only (and can even be reduced down to the size of the PSF in the extreme case of the diffusive regime), thus requiring either extremely long acquisition times or drastically reduced FoV. The high-resolution capability of this multi-angle approach (associated to recently proposed integration algorithms that enable speckle field reconstruction [48]) could be especially relevant to measure highly perturbed wavefronts. Ultimately, the method has the potential to allow the single-shot characterization of the transmission matrix [49].

When integrated into a full AO system (i.e. including hardware wavefront compensation), a single multi-angle WFS could provide instantaneous tomographic-like characterization of a large aberrating volume while significantly reducing the complexity of multi-conjugate AO systems. Given the complexity associated with using multiple compensators, an interesting compensation strategy entails using a single compensator in the pupil to correct the average aberrations in the FoV, followed by a deconvolution process in each patch [20]. Another promising strategy involves the use of a multi-angle (or "multi-pupil") compensator in the pupil [7] to correct several (2D or 3D) isoplanatic patches in real-time.

Beyond microscopy, the main fields where this parallel approach should prove most valuable are arguably those in which AO has become indispensable: astronomy and ophthalmology. In both cases, aberrations vary rapidly and isoplanatic patches are relatively small. While we demonstrated multi-angle WFS using fluorescent GSs, it can indeed be applied to other types of GS [50,51] provided that they are mutually incoherent.

While diffuser-based WFS promise cost-efficiency for future applications, the associated speckle patterns have the drawback of spreading the energy over many pixels. This is non-ideal when dealing with low light levels, especially when a large number of wavefronts is multiplexed. The use of an optimized phase mask able to generate orthogonal patterns with energy concentrated on a few pixels, such as a random contour [52,53], offers an interesting opportunity to improve signal-to-noise ratios.

Besides a wide range of applications in adaptive optics, we envision that multi-angle wavefront sensing should open new possibilities in optical diffraction tomography where the 3D refractive index mapping of the sample is usually obtained by sequentially measuring several wavefronts under multiple illumination angles [54,55]. Multi-angular WFS should allow multiplexing these measurements to greatly increase the temporal resolution, and even enable single-shot tomography, either in the visible or in the X-ray domain [56].

**Funding.** Société d'Accélération du Transfert de Technologies - ERGANEO (Project 520); DIM ELICIT - Region Ile de France (3-DiPSI); French Research National Agency (ANR-20-CE42-0006))

**Acknowledgements.** The authors thank Baptiste Blochet for stimulating discussions.

**Disclosures.** The authors declare the following competing financial interests: P.B, M.G, T.W, and G.T have filed a patent application related to angularly multiplexed wavefront sensing (US20220099499A1).

**Data availability.** The data that support the plots within this paper and other findings of this study are available from the corresponding author upon reasonable request.

**Supplemental document.** See Supplement 1 for supporting content.

# Multiplexed wavefront sensing with a thin diffuser: supplemental document

TENGFEI WU,[1,2] MARC GUILLON,[2,3] GILLES TESSIER,[1] PASCAL BERTO[1,2]*

[1]*Sorbonne Université, CNRS, INSERM, Institut de la Vision, 17 Rue Moreau, 75012 Paris, France*
[2]*Université de Paris, SPPIN – Saints-Pères Paris Institute for the Neurosciences, CNRS, 75006 Paris, France*
[3]*Institut Universitaire de France (IUF), Paris, France*
*\*Corresponding author: pascal.berto@u-paris.fr*

## S1. MULTIPLEXED MEASUREMENT WITH A PERIODIC PATTERN: AMBIGUITY ISSUE

This section aims at illustrating the main advantages of speckle patterns, as compared to periodic patterns, for the angular multiplexing of incoherent wavefronts. To this aim, we first consider a periodic pattern such as those generated by the micro-lens arrays traditionally used in Shack Hartmann wavefront sensors (WFS). In this example, we modeled a 6x6 micro-lens array, each with a micro-pupil size $p$=150μm and a focal length $f$=4.5mm. Classically, the conditions avoiding ambiguous overlap with neighboring macropixels gives a maximum dynamic range of $p/2f \approx 0.0167 \approx 1°$ for this WFS. Here, we consider that three wavefronts are multiplexed in a single model experiment [see Fig. S1(a)]: A first wavefront (red) propagates along the optical axis, while the other two wavefronts (green and blue) are tilted by an angle $\langle \alpha \rangle$ =1.6° larger than the maximum range 1° along the x and y axes. In addition to this tilt, each wavefront is formed by a linear combination of various Zernike modes with random amplitudes. The amount of tilt also contributes to inducing a translation of the pattern that exceeds the dynamic range of the macropixel, leading to a translation overlapping adjacent macropixels. In the multiplexed pattern of Fig. S1(b), we can obviously see in the macropixel marked with a dashed circle that only the red wavefront information is left, while the other two (blue and green) are translated by the tilt to regions ascribed to other macropixels. We can also observe on the zoomed insets in Fig. S1(b) that periodicity induces an ambiguity: the red and blue dots present in the same macropixel as the red one should actually be ascribed to other regions. This problem can be circumvented by designing the Shack-Hartmann WFS in such a way that the spots do not shift more than one lenslet radius, i.e. in such a way than $\langle \alpha \rangle < p/2f$. This can be achieved by increasing respectively the size of the micro-lenses $p$ or decreasing $f$, but this comes at the price of a highly reduced phase sensitivity or spatial resolution, respectively.

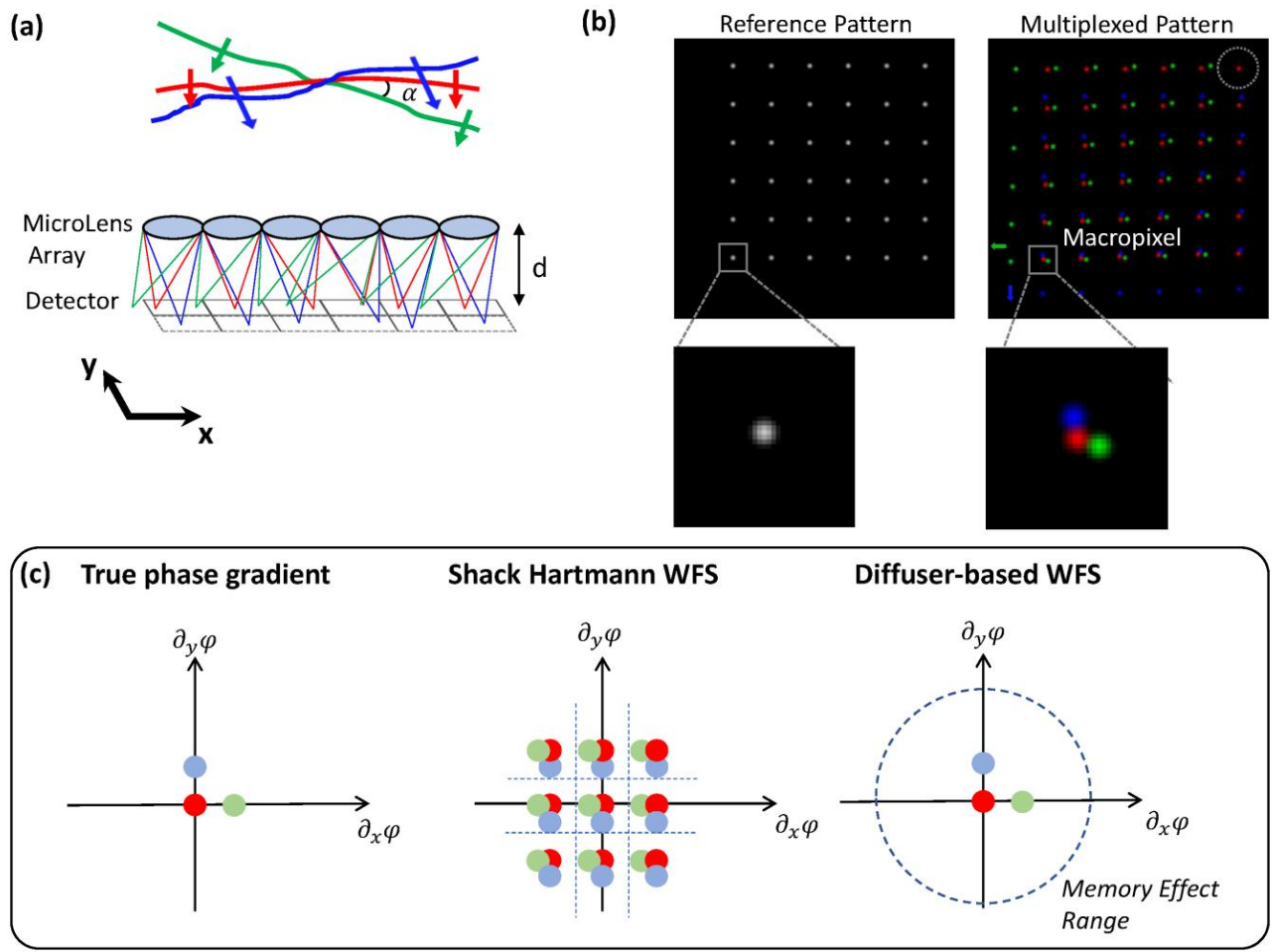

**Fig. S1.** Ambiguity problem of multiplexed measurement with periodic pattern. (a) Schematic of simultaneously sensing three wavefronts with a micro-lens array. The large tilt angle between the wavefronts induces translation, leading the local focus to shift across macropixels. (b) A numerical simulation is performed to illustrate the ambiguity issue of the multiplexed measurement with micro-lens array. The zoomed macropixel contains the wavefront information from the adjacent pixels because of the translation, while it is not distinguishable to extract the corresponding local phase gradient. (c) Representation of the ambiguity problem in the phase gradient space for the Shack Hartmann WFS and comparison with the diffuser based WFS.

In contrast, the random nature of speckle pattern implies that a speckle pattern, encoding different tilted wavefronts, are orthogonal in each macropixel of the WFS, as illustrated in Fig. 1 of the main text. This property allows multiplexed wavefronts to be distinguished even for large angular distances between tilted wavefronts, and without imposing any constraints on the dynamic range of the WFS. Figure S1(c) shows a comparison of these two wavefronts sensing methods in the phase gradient space. Note however that diffuser-based wavefront sensing imposes a constraint on the maximum detectable angle: the propagation directions of the wavefronts need to be in the range of the diffuser memory effect to allow speckle pattern recognition by intercorrelation. In practice, however, the use of a thin diffuser with a near-infinite memory effect range (i.e. a surface diffuser) alleviates this constraint.

## S2. WAVEFRONT REASSIGNMENT CRITERION

We demonstrated that a thin diffuser can be used to simultaneously measure multiple wavefronts with a single acquisition by exploiting the orthogonality of speckle patterns. For a thin-diffuser based WFS [1], a single wavefront can be unambiguously retrieved using a digital-image-correlation (DIC) algorithm: it corresponds to the unique maximum of the local cross-correlation between each macro pixel $M_A$ of the distorted speckle and the reference speckle $R$. This cross-correlation peak $M \star R$, however, is not Dirac-like, as it is broadened by both the speckle grain size and the wavefront gradient distribution.

When $N$ wavefronts are measured with a thin diffuser exhibiting a large memory effect [2,3], the multiplexed speckle pattern is an intensity superposition of $N$ replicas of the reference speckle patterns, each having undergone intensity and geometrical transformations caused by the distortion of each wavefront. Therefore, the cross-correlation $M \star R$ contains $N$ peaks instead of a single one. In the phase gradient space, these peaks are separated by a distance corresponding to the angular distance $\langle \alpha \rangle$ separating the wavefronts. Strongly distorted wavefronts, i.e. with strong local wavefront gradients, separated by a small angular distance $\langle \alpha \rangle$, can therefore lead to situations where peak overlap forbids an unambiguous separation. Note that in these conditions, similar peaks overlap also occurs with Shack-Hartman WFS.

In order to distinguish two wavefronts in a multiplexed speckle, their angular separation $\langle \alpha \rangle$ must be larger than the lateral width of the cross-correlation peaks (which is driven by both the speckle grain size and the phase gradient distribution, excluding the tilt $\langle \alpha \rangle$). Here, we numerically investigate this condition.

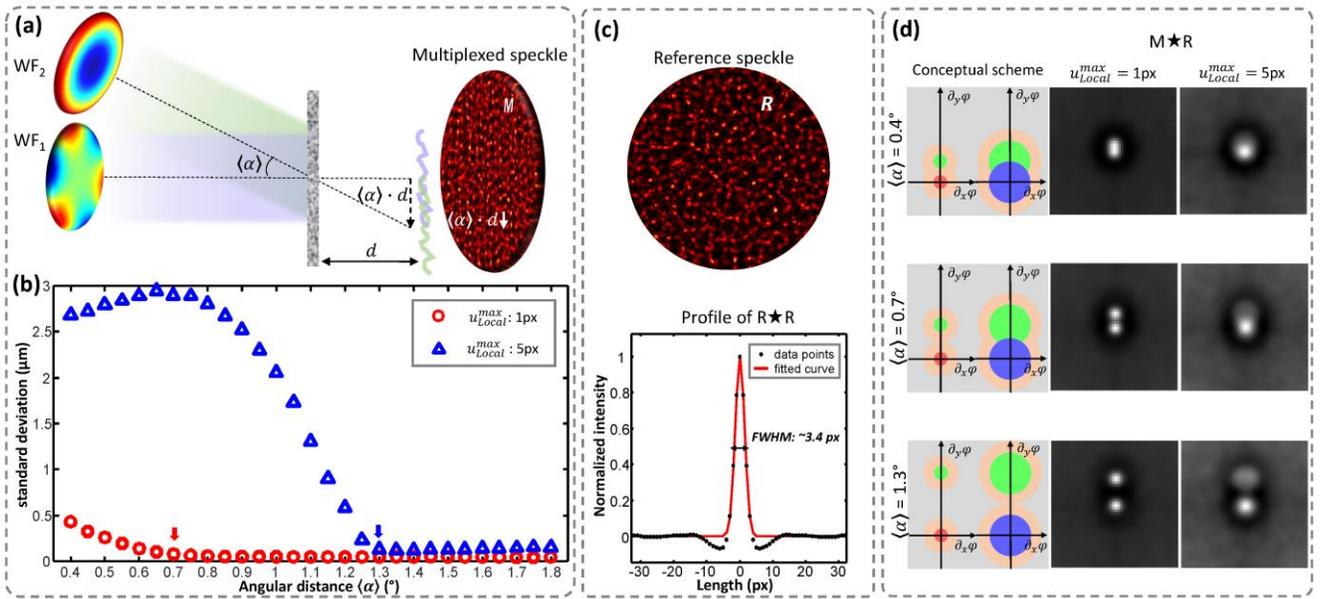

**Fig. S2. Numerical simulation: criterion to reassign wavefronts in a multiplexed speckle Pattern M.** (a) Two wavefronts corresponding to pure Zernike Polynomials with an average angular distance $\langle \alpha \rangle$ are multiplexed using a thin diffuser located at a distance $d$ from the camera. The resulting speckle grains displacements $\boldsymbol{u}_j = \langle \boldsymbol{u} \rangle_j + \boldsymbol{u}_{Local,j}$ can be described as the sum of an average displacement $\langle \boldsymbol{u} \rangle_j = \langle \alpha \rangle_j \cdot d$ and local displacements $\boldsymbol{u}_{Local,j}$ corresponding to the local phase gradients (excluding the tilt, thus only related to the distortion induced by the Zernike polynomials). The two wavefronts are reconstructed from the multiplexed speckle pattern $M$ and compared to the wavefronts used as input. (b) Evolution of the reconstruction error (standard deviation) plotted against the angular distance $\langle \alpha \rangle$. For both wavefronts, we apply an amplitude coefficient to the Zernike polynomials which results in a chosen maximum local displacement $\boldsymbol{u}_{Local}^{max} = \boldsymbol{u}_{Local,1}^{max} = \boldsymbol{u}_{Local,2}^{max} = 1px$ (*red cicle*) or 5px (blue triangle). The inflection points indicated by arrows correspond to the minimum angular distances $\langle \alpha \rangle$ allowing wavefront isolation. (c) the reference speckle $R$ and its autocorrelation, from which the speckle grain size is measured. (d) Conceptual description of the cross-correlation peaks in the presence of various phase gradients (radii of the green, blue and red discs) and speckle-induced broadening (orange discs). Broad peaks separated by a small angle $\langle \alpha \rangle$ (top) overlap, but can be separated (bottom) for larger angular distances.

To model the 1° holographic diffuser (Edmund Optics) used as Hartmann mask, we first characterize its physical parameters by conjugating the diffuser surface to a commercial high-resolution WFS (PHASICS, SID4). Based on the acquired quantitative phase map, we extract a phase correlation width $w_{FWHM}$=46.1μm and a phase standard deviation $\bar{\delta}$ =0.156μm. Using these parameters, we numerically generate a pseudo-random phase mask with realistic statistical properties, as explained and validated in [4].

In order to investigate the condition to properly distinguish wavefronts in a multiplexed measurement, we performed the numerical simulation shown in Fig. S2(a) and S2(b). Two wavefronts (Zernike polynomials: WF$_1$=Vertical trefoil $Z_3^{-3}$ and WF$_2$=Defocus $Z_2^0$) are used here. WF$_1$ is assumed to propagate along the optical axis, and WF$_2$, along the angular direction $\langle \alpha \rangle$ which varies from 0.4° to 1.8° with steps of 0.05°. The resulting speckle grains displacements $\boldsymbol{u}_j = \langle \boldsymbol{u} \rangle_j + \boldsymbol{u}_{Local,j}$ can be described as the sum of an average displacement $\langle \boldsymbol{u} \rangle_j = \langle \alpha \rangle_j \cdot d$ and local displacements $\boldsymbol{u}_{Local,j}$ corresponding to the local phase gradients (excluding the tilt, thus only related to the distortion induced by the Zernike polynomials). For both wavefronts, we apply an amplitude coefficient to the Zernike polynomials which results in a chosen maximum local displacement $\boldsymbol{u}_{Local}$ (namely, $\boldsymbol{u}_{Local,1}^{max} = \boldsymbol{u}_{Local,2}^{max} = 1$ or 5 px). The propagation of the optical field between the diffuser and the camera sensor is calculated using Fresnel diffraction, for a diffuser-camera distance $d$=3mm and a pixel size of 6.5µm. The angular distance $\langle \alpha \rangle$ induces a lateral shift $\langle \alpha \rangle \cdot d$ between the two speckle patterns, as shown in Fig. S2(a). The speckle grain size is measured in Fig. S2(c) as the FWHM of a Gaussian fit on the autocorrelation of the reference speckle pattern $R$, which yields a size of 3.4 px. Using direct DIC, we evaluate wavefront WF$_1$.

To evaluate the quality of the wavefront reconstruction, we calculate the standard deviation of the difference between the reconstructed and the input wavefront for various values of the angular separation $\langle \alpha \rangle$. Fig. S2(b) shows this error calculated for $\boldsymbol{u}_{Local,1}^{max} = \boldsymbol{u}_{Local,2}^{max}$ =1px and 5px (angular broadening applied to both WF$_1$ and WF$_2$). As $\langle \alpha \rangle$ increases, the measurement error converges to a minimum in both cases. The inflection point, however, depends on the value of $\boldsymbol{u}_{Local}^{max}$: a strong WF gradient, corresponding to a broad distribution ($\boldsymbol{u}_{Local}^{max}$ = 5 px), requires larger separations ($\langle \alpha \rangle$=1.3°) to be properly reconstructed. Fig. S2(d) shows the conceptual scheme in phase gradient space for the two cases of $\boldsymbol{u}_{Local}^{max}$. The red disk (1 px) and blue disk (5 px) respectively represent the phase gradient pattern of fixed WF$_1$. The green disk represents WF$_2$, globally tilted along the y axis with a value of $\langle \alpha \rangle \cdot d$. These phase gradient disks are additionally broadened by a kernel determined by the speckle grain size (orange). For $\langle \alpha \rangle = 0.4°$, which corresponds to $\langle \alpha \rangle \cdot d \approx 3.1$ px, the peaks cannot be resolved since the speckle grain size is around 3.4 px. For $\langle \alpha \rangle = 0.7°$, the two peaks corresponding to $\boldsymbol{u}_{Local}^{max}$ = 1 px are well separated, but the $\boldsymbol{u}_{Local}^{max}$ = 5 px peaks are not. Finally, for $\langle \alpha \rangle$ =1.3°, all peaks are well separated, thus allowing proper multiplexed measurements of the wavefronts.

## S3. ITERATIVE DIGITAL-IMAGE-CORRELATION PROCESS

This section discusses (S3.1) the limits of the single-iteration (or direct) DIC algorithm used to recover the wavefronts from the multiplexed speckle image and (S3.2) provides a detailed description of the iterative DIC algorithm as well as (S3.3) an evaluation of its performances.

### S3.1 Limits of the Direct DIC algorithm

As explained in the main text, the multiplexed speckle pattern $M$ observed in the camera plane is the intensity superposition of the $N$ speckle patterns $S_j$ related to each Guide Star (GS): $M(r) = \sum_{j=1}^{N} S_j(r)$. The speckle contrast $C \propto 1/\sqrt{N}$ thus decreases when increasing the number $N$ of multiplexed wavefronts. This induces errors when estimating the speckle grains displacement $u_j$ (i.e phase gradients) and thus, ultimately affects the fidelity of the wavefront reconstruction. This section aims at studying this effect, in particular the influence of both the phase pixel size and the number of multiplexed wavefronts. To this aim, we numerically consider a 4x4 array of GS described in Fig. S3(a). The angular distance between neighboring GSs is fixed to: $\alpha = 0.8°$. The tilted wavefronts emitted by each of these GSs are distorted by a vertical trefoil aberration with a Peak-to-Valley amplitude of 8λ. The wavefront obtained when only one GS is activated ($N$=1, no multiplexing) is used as ground truth, and compared to the wavefronts obtained when increasing the number of activated GSs, from $N$=2 to 16. Figure S3(c) shows the normalized wavefront reconstruction error as a function of both the number of multiplexed wavefronts and the size $n$ of the phase pixel (where $n$ corresponds to the macropixel formed by $n \times n$ camera pixels, with $n$ ranging from 7 to 45 pixels). For small phase pixel sizes, we can see that the reconstruction error increases quickly with the number of multiplexed wavefronts. This is due to the fact that the cross-correlation between a macropixel $M_A$ and the reference $R$ (see Fig. S3(b), left) is subject to a higher level of noise ($\sigma \propto \sqrt{N}$), which degrades the peak centroid estimation. Interestingly, the signal-to-noise ratio of the cross-correlation map $M_A \star R$ can be increased by increasing the macropixel size $M_A$, as shown in Fig. S3(b). As shown in Fig. S3(c) and (d), the reconstruction error decreases rapidly when larger macropixels (or phase pixels) are used.

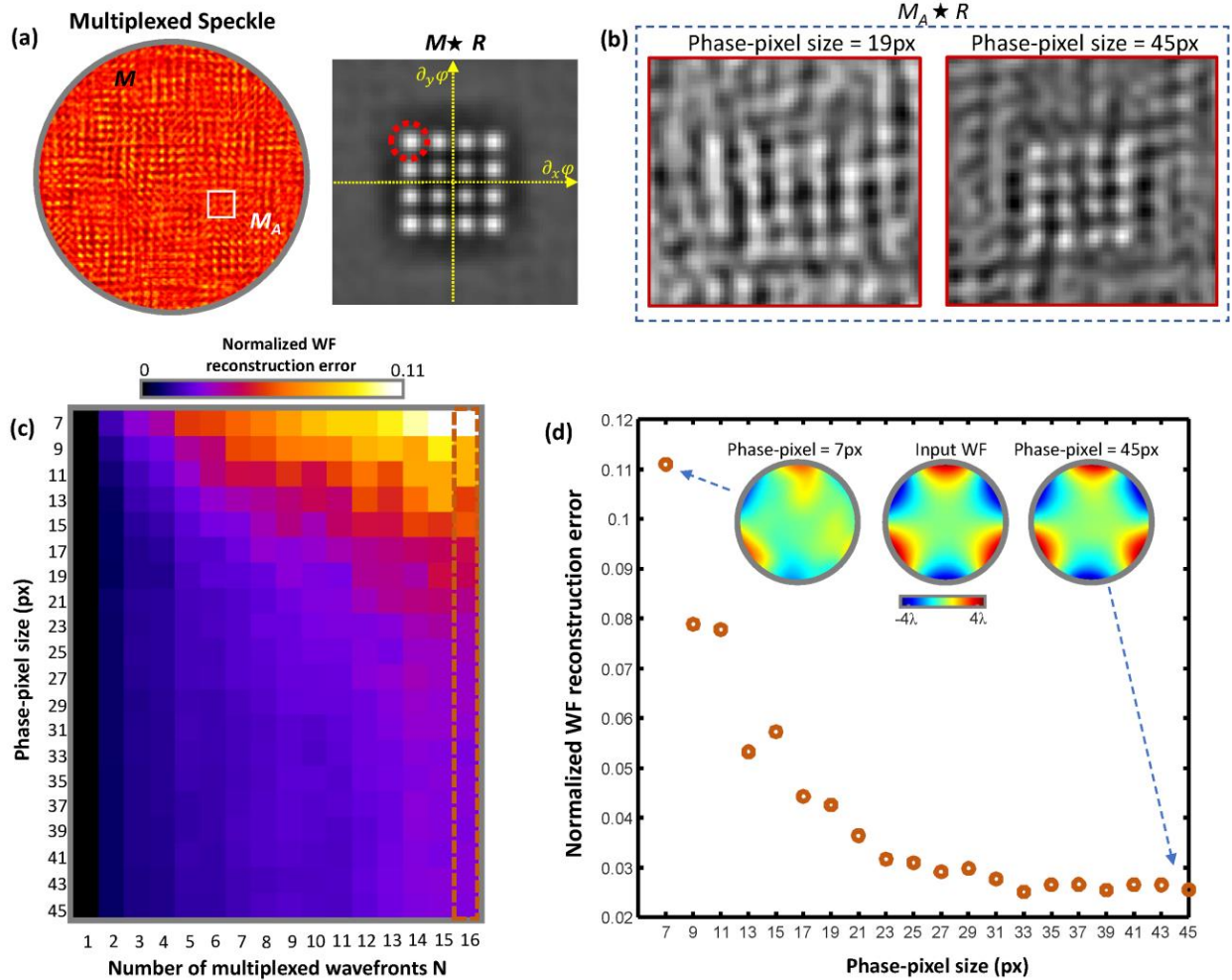

**Fig. S3.** Numerical simulation results show the reconstruction error of the "direct DIC algorithm" depending on the phase pixel size and number of multiplexed wavefronts. (a) Right: distribution of the simulated GS array in the phase gradient space (or k-space). Left: multiplexed speckle pattern from $N$=16 wavefronts. (b) Cross-correlation map between the macropixel $M_A$ and the reference $R$ when considering two macropixel sizes: $n \times n$=19x19 and 45x45 pixels. The SNR can be increased by increasing the macropixel size. (c) Performances of the direct DIC algorithm depending on the number of multiplexed wavefronts $N$ and the phase pixel size $n$. The reconstruction error increases for large numbers of wavefronts and smaller pixel sizes. (c) Evolution of the reconstruction error with the phase pixel size (considering $N$=16 multiplexed wavefronts).

### S3.2 Description of the Iterative DIC Process

The previous section has demonstrated that in order to maintain high reconstruction fidelity when multiplexing a large number of wavefronts, it is necessary to use large macropixel sizes. However, using large macropixels limits the ability to perform high-resolution wavefront sensing. To overcome this limitation, we propose to use the DIC algorithm in an iterative manner to enable the use of small macropixel sizes without sacrificing reconstruction accuracy. The proposed strategy consists in isolating iteratively each speckle pattern $S_j$ before identifying the final speckle grains displacement map. An individual distorted speckle map $S_k$ related to the $k^{th}$ wavefront can indeed be isolated from the multiplexed speckle $M$ by iteratively subtracting an estimation of the other distorted speckle patterns $S_j$ ($j \neq k$): $S_k(r) = M(r) - \sum_{j \neq k}^{N} S_j(r)$. To this aim, the distorted speckle patterns $S_j$ are built at each step by applying the estimated maps of the speckle grains displacement $u_j$ and the intensity $I_j$ to the reference $R$:

$$S_j(r) = I_j(r) R[r - u_j(r)] \tag{1}$$

The contrasted speckle patterns $S_k$ are then used in the DIC process to iteratively extract a better estimation of $u_k$ and $I_k$, and reconstruct each wavefront with a higher accuracy.

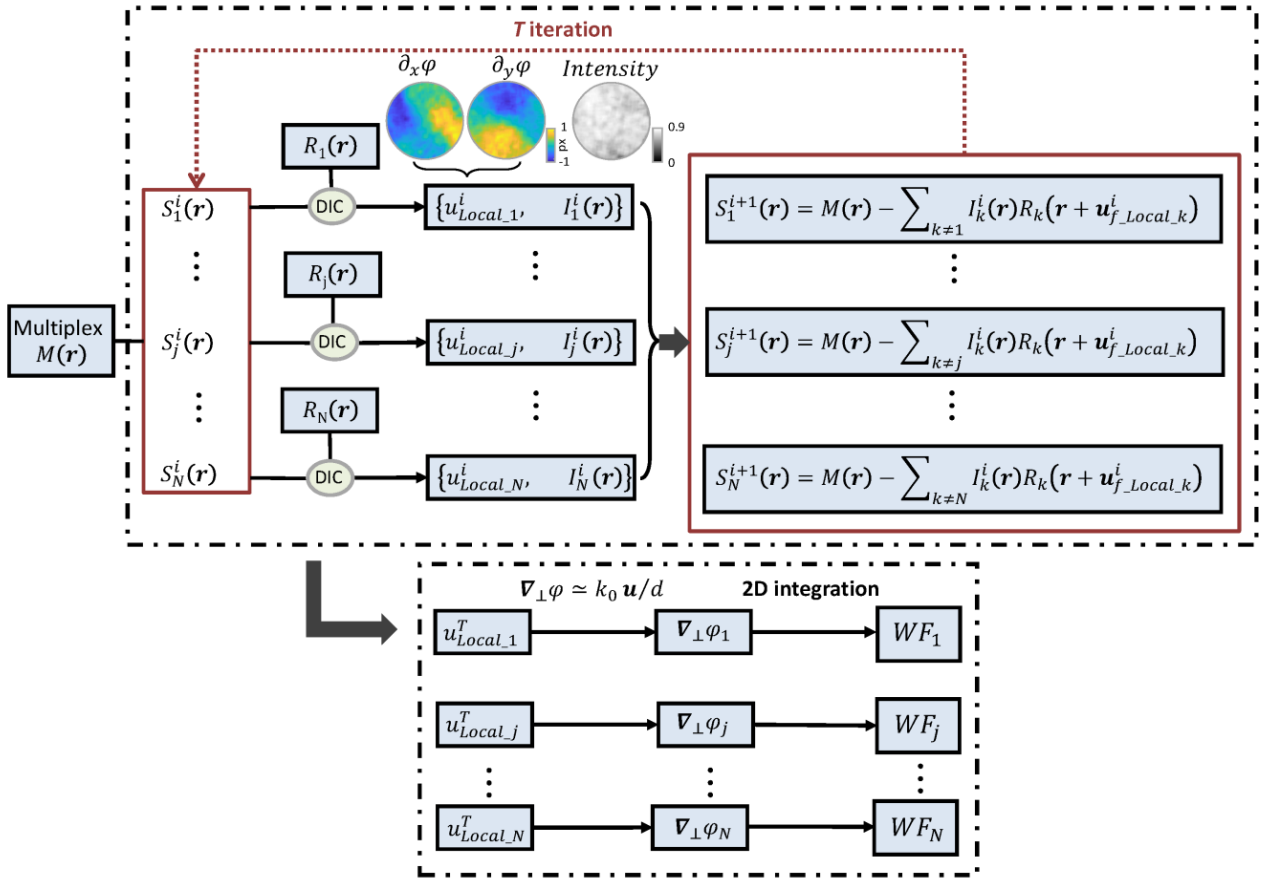

**Fig. S4.** Flow diagram of the iterative DIC process

A detailed flow diagram of the iterative DIC process is shown in Fig. S4.
1. In a preliminary step (not shown in Fig. S4.), the global cross correlation between the multiplexed and the reference speckles, $M \star R$ is calculated (i) to extract the number $N$ of wavefronts and (ii) to estimate, by measuring the centroid of each correlation peak, the global speckle displacement $\langle u \rangle_j = \langle \alpha \rangle_j \cdot d$ associated to their global tilt $\langle \alpha \rangle_j$. For each GS, a reference speckle pattern $R_i$, is then generated by simply translating the original reference speckle pattern $R$: $R_i = R(r - \langle u \rangle_j)$.
2. For the first iteration $i=1$, the multiplexed speckle $M(r)$ is used as the "initial isolated speckle $S_j^i(r)$" for each wavefront.

3. The DIC algorithm is then applied on $S_j^i(r)$ and $R_i$ to measure the intensities $I_j^i(r)$ and the local displacement maps $u_{Local\_j}^i$ defined as $u_j^i = \langle u \rangle_j + u_{Local\_j}^i$.
4. *Optional spatial filtering step (not shown in Fig. S4)*: to reduce the high frequency noise in the displacement map, we spatially filter the $u_{Local\_j}^i$ map associated to the gradient map in the following way: we first obtain an initial wavefront from the $u_{Local\_k}^i$ maps, and perform a Zernike decomposition to remove very high order Zernike modes (modes above #67 modes are removed to preserve the full wavefront information). We then calculate the gradient of this spatially filtered wavefront and retrieve the filtered displacement map $u_{f\_Local\_j}^i$.
5. The information of $\{u_{f\_Local\_j}^i, I_j^i(r)\}$ is used to register and weigh the reference speckle pattern $R$ in order to estimate the distorted speckle pattern corresponding to each GS. Each isolated speckle $S_k^{i+1}(r)$ can then be obtained by subtracting the others from the multiplexed speckle: $S_j^{i+1}(r) = M(r) - \sum_{k \neq j} I_k^i(r) R_k\left(r + u_{f\_Local\_k}^i\right)$
6. $S_k^{i+1}(r)$ is used in the next iteration where steps 3,4 and 5 are repeated. The final isolated speckle patterns are obtained after T iterations and used to extract the final phase gradients based on the knowledge of the distance $d$ between the thin diffuser and the camera sensor.
7. A 2D integration algorithm finally allows to reconstruct each wavefront $WF_N$ from the phase gradient [5].

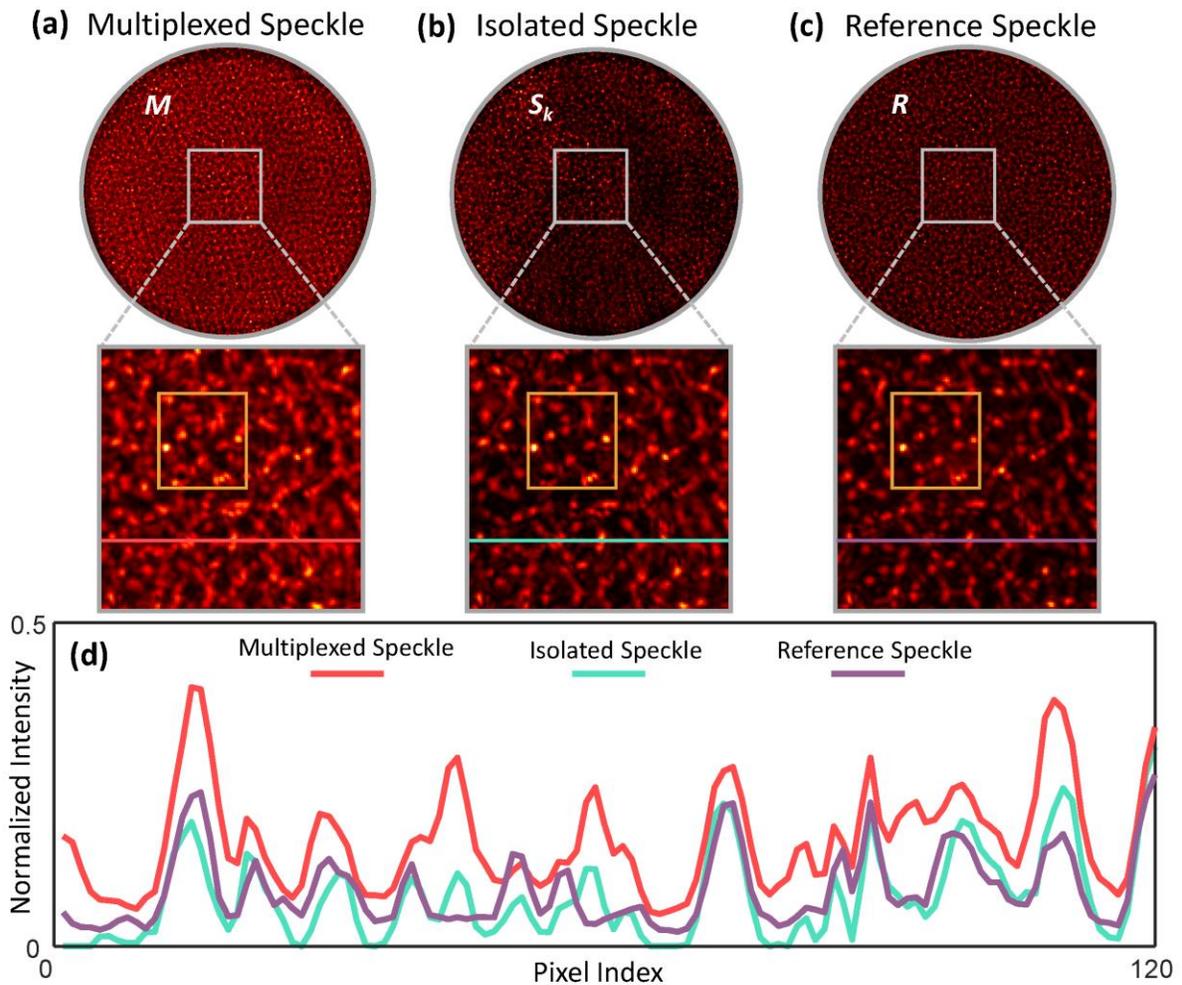

**Fig. S5. Illustration of the speckle pattern isolation from the iterative DIC process.** (a) multiplexed speckle, (b) isolated speckle and (c) reference speckle. The speckle grains contrast of the three speckle patterns are compared in zoomed regions (bottom) and corresponding profiles are shown in (d).

In order to illustrate the speckle isolation process, Fig. S5(a), (b) and (c) compare the multiplexed speckle $M$ encoded with $N=3$ wavefronts, the isolated speckle $S_j$ from the aforementioned iterative process and the reference speckle $R$. We can clearly notice that the contrast of the isolated speckle pattern is increased as compared to the multiplexed one and becomes comparable with the one of the reference. The speckle grains in the isolated speckle are more resolvable and contrasted, which is useful to precisely estimate the speckle grains displacement using DIC. Figure S5(d) further compares the speckle profiles.

### S3.3 Performances of the Iterative DIC Process

Numerically, we further study in Fig. S6 the relation between the number of multiplexed wavefronts and the number of iterations which are necessary to reconstruct them. Here, we consider the same 4x4 GS array described in Fig. S3(a). The angular distance between neighboring GSs is again fixed to: $\alpha = 0.8°$. We then sequentially increase the number of activated GS in the array, reconstruct the wavefront, and compare it to a ground truth to estimate the reconstruction error. The ground truth we consider here is the wavefront used as input in the simulation.

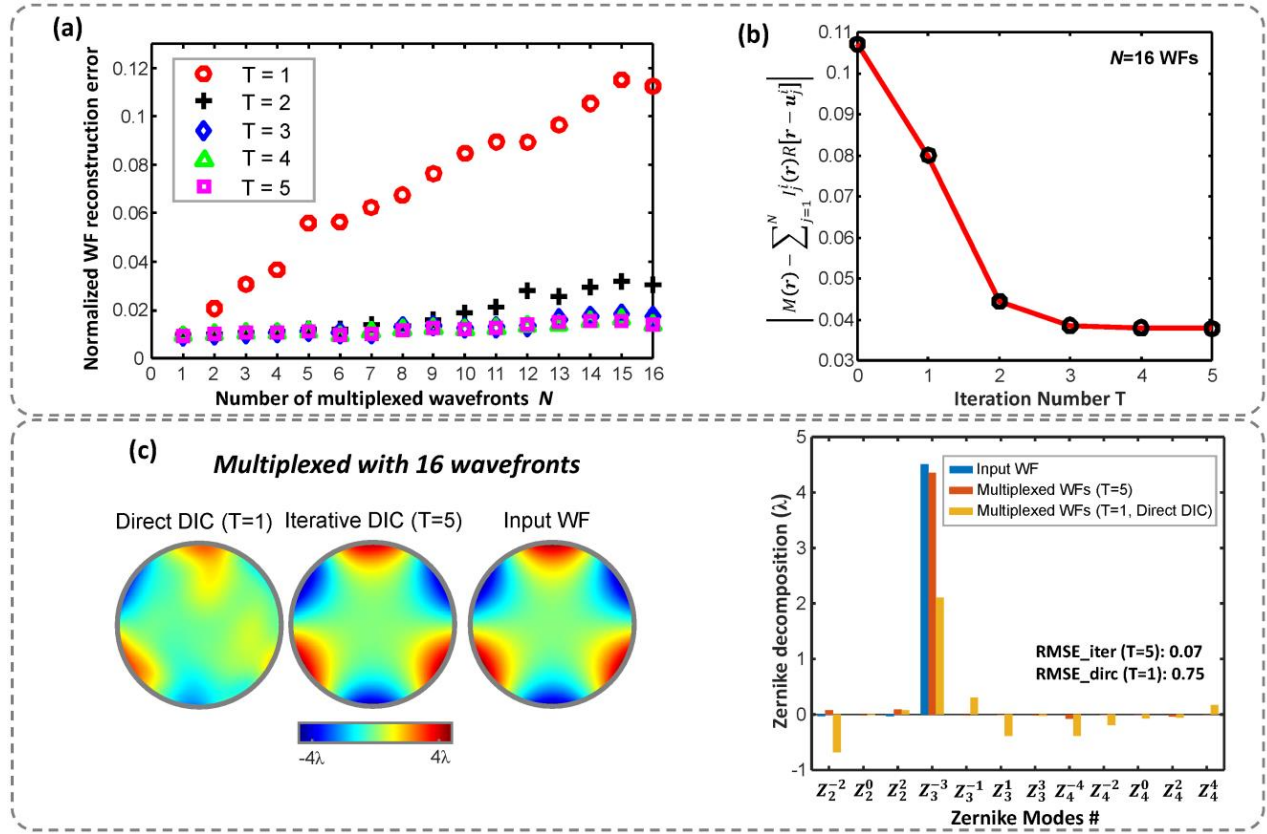

**Fig. S6.** Numerical simulation results showing the gain provided by the iterative DIC process for reconstructing multiplexed wavefronts with high resolution (considering a phase pixel size: 7x7 px). (a) Performance of the iterative DIC algorithm depending on the number of multiplexed wavefronts. T=1 corresponds to the direct DIC approach. We can see that when increasing the number of multiplexed wavefronts, the required number of iterations increases but allows high reduction of the reconstruction noise. (b) Convergence of the algorithm for N=16 multiplexed wavefronts. (c) Comparison of the wavefronts respectively reconstructed by direct DIC and iterative DIC (considering N=16 multiplexed wavefronts).

To illustrate the potential of the method to reach high-resolution wavefront sensing, we considered a 7x7 phase macropixel. As expected, Fig. S6(a) shows that when using a single iteration of the DIC algorithm (T=1 or "direct" DIC, red circles), the reconstruction error increases with the number N of multiplexed wavefronts. Interestingly, we can see that the reconstruction error decreases sharply with subsequent iterations of the DIC process. Furthermore, we notice that the number of iterations required to achieve an optimal reconstruction increases with the number of multiplexed wavefronts (typ: T=2, for N=5; T=3 for N=11). Finally, Fig. 6(c) compares the wavefront measurement from direct DIC and iterative DIC when N=16 GSs are multiplexed and shows the Zernike decomposition of the results. We can clearly see that the iterative DIC algorithm allows to reconstruct a large number of wavefronts (N=16) while preserving small macropixel (7x7 px) which enable high-resolution multiplexed wavefront sensing.

## S4. DETAILED DESCRIPTION OF THE SETUP

***Selective GSs excitation and wavefront measurement -*** The setup used to selectively excite GSs and measure their wavefronts is shown in Fig. S7(a). A 532nm laser beam is modulated by a phase-only spatial light modulator (SLM, X13138-01, Hamamatsu) conjugated to the back focal plane of the microscope objective (LUCPlanFLN, NA=0.45, x20, Olympus). A computer-generated hologram can be loaded in real time on the SLM to generate multiple focal spots in the sample plane in order to simultaneously excite fluorescent beads (or GSs) of interest [see Fig. S7(a)]. The phase-only hologram is calculated using a standard Gerchberg-Saxton algorithm. The excitation light is filtered out by a dichroic mirror (NFD01-532, Semrock) and a notch filter ($\lambda$=533 ± 2 nm, Thorlabs) to collect the emitted fluorescence. As shown in the main text, the multiplexed WFS, conjugated to the microscope objective pupil, consists of a thin diffuser (1° scattering angle holographic diffuser, Edmund Optics) and an sCMOS camera (Zyla 5.5, Andor).

**Widefield excitation and imaging** - On the collection path, a flip mirror (FM) is used to switch towards an imaging path [see Fig. S7(b)]. A sCMOS camera (Panda 4.2, PCO) is conjugated to the sample plane by a 4f system in order to directly image the fluorescent sample. In this configuration, a wide-field excitation can be set by applying the appropriate hologram on the SLM.

***Sample and isoplanetic patch characterization -*** the sample consists of 1μm-diameter fluorescent beads randomly distributed on a glass slide (Orange 540/560, Thermo Fisher). A surface diffuser (1° Holographic Diffusers, Edmund Optics) is positioned $z$=150 μm away from the sample to act as the aberrating medium inducing spatially varying pupil aberrations [see Fig. S7(c)]. In order to properly select GSs located in different isoplanatic patches, the size of the patch needs to be estimated. To do so, we sequentially acquired a stack of speckle patterns, each of which contains a wavefront from a single excited GS. The location of each GS in this stack was also recorded and the corresponding wavefronts were calculated. Figure S7(c) shows the result of the cross-correlation coefficients between wavefronts as a function of their relative distances. A Gaussian model is used to fit the measurements and its half-width-half-maximum, i.e. the radius of the isoplanatic patch in this configuration, is estimated as $R_{patch}$≈ 55μm.

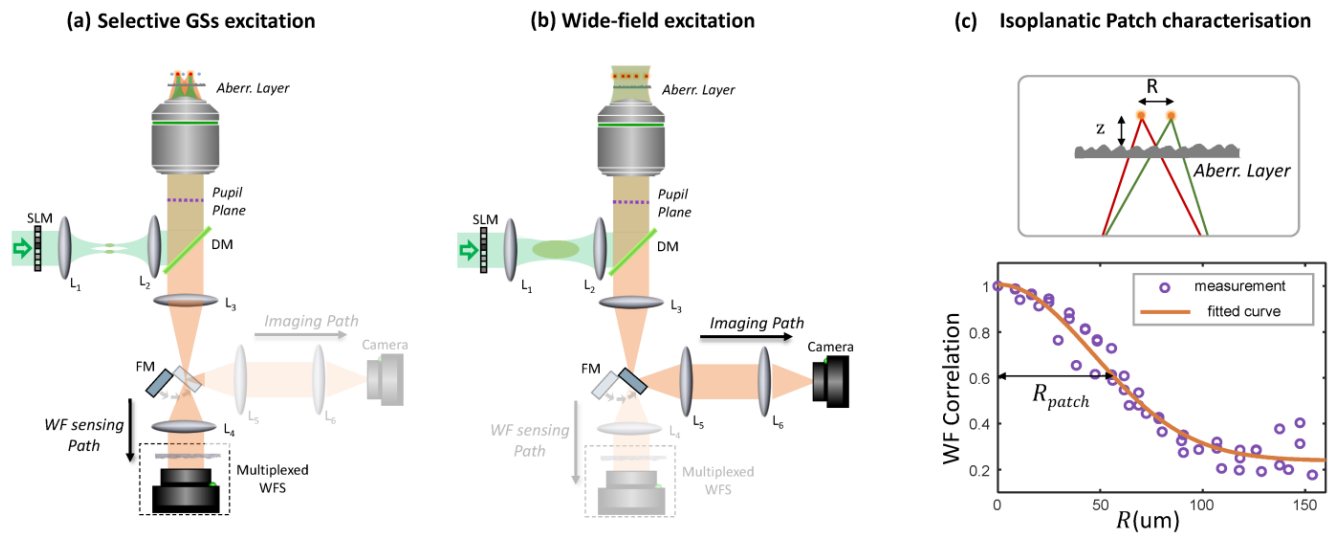

**Fig. S7. Experimental setup.** (a) A 532nm laser beam is shaped by a phase-only SLM, conjugated to the pupil plane of the microscope objective, in order to excite one or several fluorescent beads in the FoV. On the collection path, the multiplexed diffuser-based WFS is conjugated to the pupil plane to characterize spatially varying pupil aberrations in a single shot. A flip mirror can redirect fluorescence towards (b) an imaging path where a camera is conjugated to the sample plane. (c) Sample description: an aberrating layer, here a surface diffuser, is located at a distance $z$=150μm from the fluorescent beads (or GSs) to induce aberrations varying in the FoV. The radius of the isoplanatic patch ($R_{patch}$≈55μm) is estimated by measuring the wavefront decorrelation when considering different GSs in the FoV.

## S5. EXPERIMENTAL VALIDATION OF THE ITERATIVE DIC PROCESS

This section aims at comparing the experimental performances of the direct DIC and iterative DIC algorithms. Figure S8(a) shows the reference speckle $R$ and the multiplexed speckle $M$, which has much lower contrast since it associates $N=5$ multiplexed wavefronts. By calculating $M \star R$, we can identify the angular location of the $N=5$ GSs [see Fig. S8(a), right]. Each of the corresponding peaks has a different brightness, indicating that speckle pattern has a different intensity. Figure S8(b) shows the residue $\left| M(r) - \sum_{j=1}^{N} I_j(r) R[r - u_j(r)] \right|$ calculated at each step of the iterative DIC process. T=0 corresponds to the case where only the global tilt has been considered ($u_j(r) = \langle u \rangle_j$) while T=1 corresponds to the "direct" DIC case. We can see that T=2 iterations are needed to precisely estimate the $N=5$ distorted speckle patterns from which the $N$ wavefronts are reconstructed. Interestingly, this number of iteration T is precisely the one expected from numerical simulation for $N=5$ wavefronts [see Fig. S6(a)].

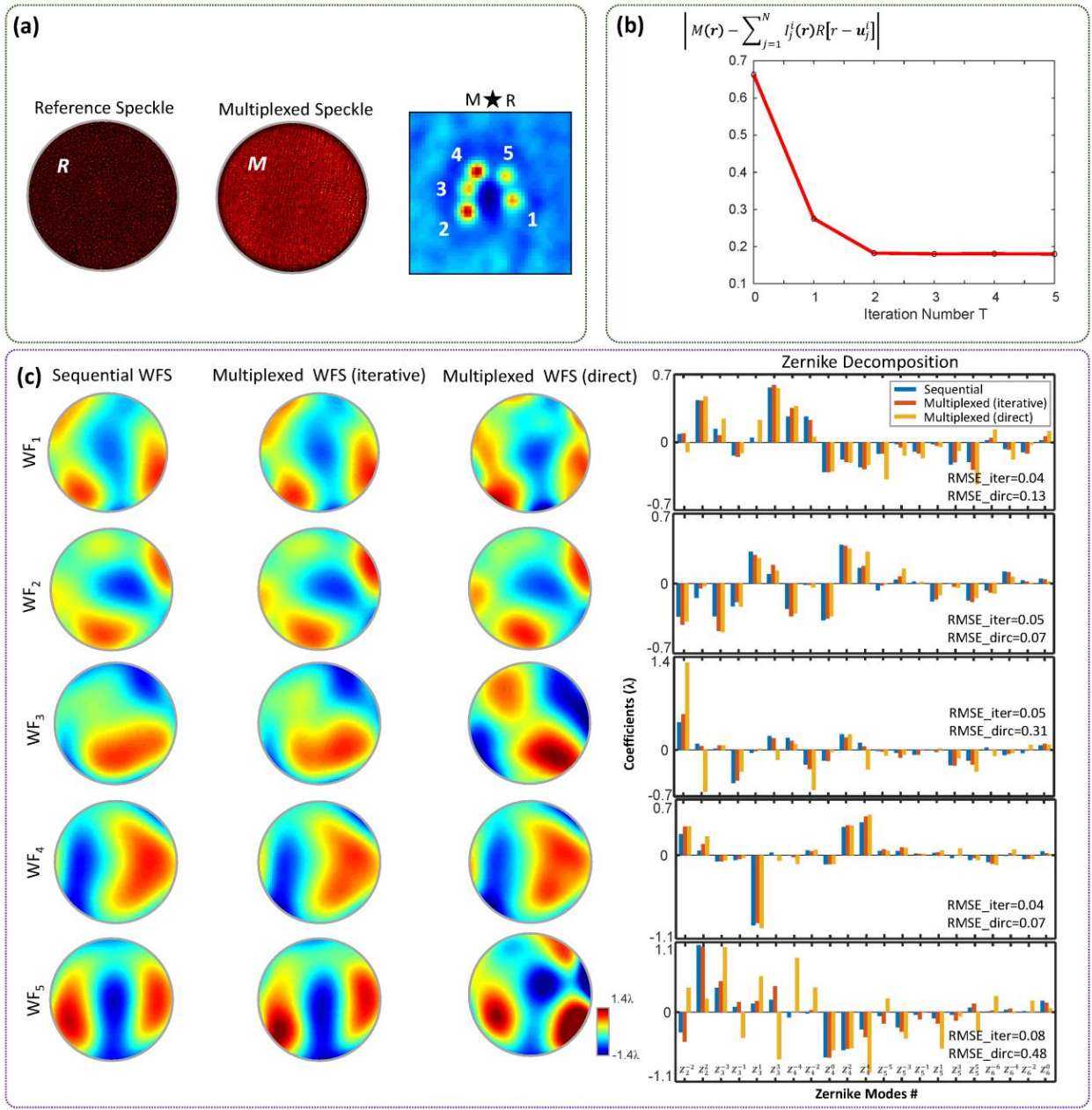

**Fig. S8. Experimental result of multiplexed measurement of five wavefronts.** (a) Multiplexed speckle pattern, reference speckle pattern and their cross-correlation, which indicated the location and the number of multiplexed GS. (b) Convergence of the iterative DIC algorithm (c) Comparison of the wavefront measurements of sequential mode, multiplexed mode with iterative DIC process and direct DIC algorithm.

The first column of Fig. S8(c) shows the 5 wavefronts measured when only one of the 5 GSs is sequentially activated. This non-multiplexed measurement can be compared to multiplexed measurements (columns 2 and 3), obtained with a single iteration (T=1 or direct) and with T=3 iterations. As can be seen qualitatively on these maps, and quantitatively in their Zernike decompositions [see Fig. S8(c), right], the results of sequential WFS and multiplexed WFS are very similar when using the iterative approach, but some differences clearly appear when considering the single-iteration procedure. The results of this direct DIC algorithm are indeed influenced by the low contrast of the multiplexed speckle, a detrimental effect which is alleviated by the iterative procedure. We also notice that since the intensity of the individual speckle patterns from $WF_2$ and $WF_4$ are higher than the others (from the cross-correlation $M \star R$, indicating a better identification in the multiplexed speckle pattern), the displacement maps have less errors than the other three wavefronts in the direct DIC algorithm, enabling better wavefront measurements.

## S6. ABERRATION CORRECTION OVER A LARGE FIELD OF VIEW WITH MULTIPLEXED MEASUREMENT

As discussed in the main text, once the pupil aberration is properly measured, it can be corrected digitally with a deconvolution process, in which a point-spread function (PSF) can be calculated from the measured pupil aberration.

In Fig. S9(a) we show a non-aberrated image of randomly-distributed fluorescent beads, when the aberrating medium is removed (Ground truth). Three beads chosen in different regions of the image are selectively excited using the setup described in Supplementary S4, to be used as GS, and the corresponding wavefronts are measured.

Due to the limited size of isoplanatic patch ($R_{patch} \approx 55\mu m$), these wavefronts, i.e. the aberrations, are statistically uncorrelated in the corresponding GS regions. The PSFs deduced from these measurements can only be used to correct aberration in the vicinity of the GS used to measure it, i.e. within the corresponding isoplanatic patch. We deconvolve sequentially the degraded image with the PSFs calculated for the three GSs and the deconvolution is performed with 15 iterations of the Richardson-Lucy algorithm (Matlab Image Processing Toolbox). With each GS as the center, we zoom in and compare the corresponding 70μm square regions.

As shown in Fig. S9(c) the resolution and contrast obtained in region ① are optimal (and similar to those of the unaberrated image) when using the PSF measured using $GS_1$. Corrections performed using the PSFs measured with $GS_2$ and $GS_3$ provide a notable improvement over the raw aberrated image, but these corrections remain inferior to what is obtained with $GS_1$. Similarly, regions ② and ③ are, as expected, best corrected when using the wavefronts measured with $GS_2$ and $GS_3$, respectively. As detailed in the main text, a proper stitching of images corrected within their own isoplanatic patch using the wavefronts measured simultaneously in these regions can provide a high quality, corrected image over a large FoV.

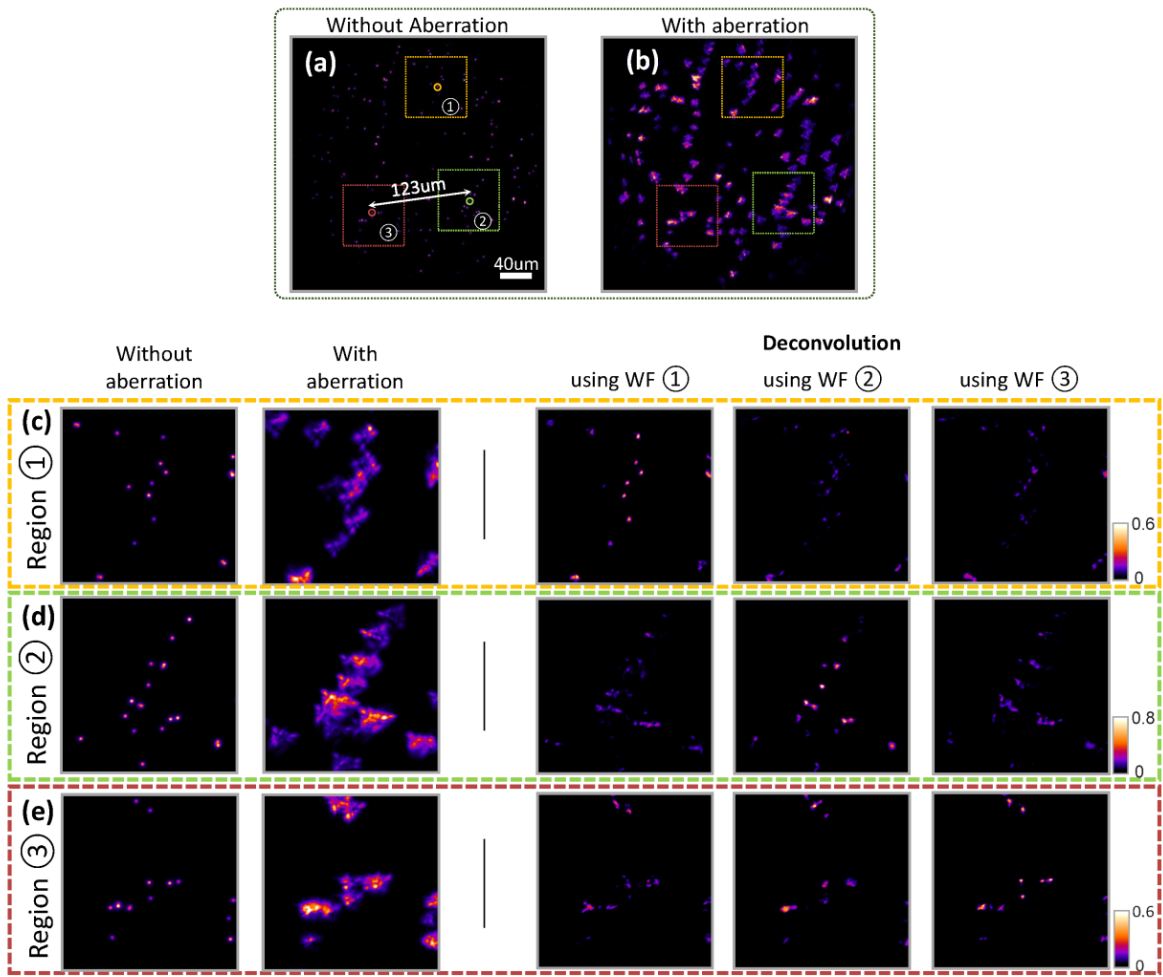

**Fig. S9. Aberration correction in different isoplanatic patches from a single shot measurement and a deconvolution process.** (a) Image of randomly distributed fluorescent beads without (a) and with (b) an aberrating layer. The fluorescent beads numbered ①-③ are selectively illuminated to be used as GS, and the three corresponding WFs are measured simultaneously. The shortest distance between these GS is around 123μm. (c)-(d) Aberration correction in different isoplanatic patches with the three PSFs, calculated from the measured pupil aberrations. Scale bar: 40μm.